\documentclass[sigconf]{acmart}

\usepackage{cleveref}
\usepackage{tabularx}
\usepackage{multirow}
\usepackage{listings}
\usepackage{glossaries}

\newcolumntype{L}[1]{>{\raggedright\let\newline\\\arraybackslash\hspace{0pt}}p{#1}}
\newcolumntype{C}[1]{>{\centering\let\newline\\\arraybackslash\hspace{0pt}}p{#1}}
\newcolumntype{R}[1]{>{\raggedleft\let\newline\\\arraybackslash\hspace{0pt}}p{#1}}

\newacronym{csd}{CSD}{continuous software development}

\AtBeginDocument{%
  \providecommand\BibTeX{{%
    \normalfont B\kern-0.5em{\scshape i\kern-0.25em b}\kern-0.8em\TeX}}}

\copyrightyear{2024}
\acmYear{2024}
\setcopyright{rightsretained}
\acmConference[ICSE-SEIP '24]{46th International Conference on Software Engineering: Software Engineering in Practice}{April 14--20, 2024}{Lisbon, Portugal}
\acmBooktitle{46th International Conference on Software Engineering: Software Engineering in Practice (ICSE-SEIP '24), April 14--20, 2024, Lisbon, Portugal}
\acmDOI{10.1145/3639477.3639736}
\acmISBN{979-8-4007-0501-4/24/04}

\begin{document}

\title{Industrial Challenges in Secure Continuous Development}

\author{Fabiola Moy\'on}
\orcid{0000-0003-0535-1371}
\email{fabiola.moyon@siemens.com}
\affiliation{%
  \institution{Siemens Technology and Technical University of Munich}
  \country{Germany}
}

\author{Florian Angermeir}
\orcid{0000-0001-7903-8236}
\email{angermeir@fortiss.org}
\affiliation{%
  \institution{fortiss}
  \country{Germany}
}

\author{Daniel Mendez}
\orcid{0000-0003-0619-6027}
\email{daniel.mendez@bth.se}
\affiliation{%
  \institution{Blekinge Institute of Technology and fortiss}
  \country{Sweden and Germany}
}

\begin{abstract}
The intersection between security and continuous software engineering has been of great interest since the early years of the agile development movement, and it remains relevant as
software development processes are more frequently guided by agility and the adoption of DevOps. Several authors have contributed studies about the framing of secure agile development and secure DevOps, motivating academic contributions to methods and practices, but also discussions around benefits and challenges. Especially the challenges captured also our interest since, for the last few years, we are conducting research on secure continuous software engineering from a more applied, practical perspective with the overarching aim to introduce solutions that can be adopted at scale.
The short positioning at hands summarizes a relevant part of our endeavors in which we validated challenges with several practitioners of different roles. More than framing a set of challenges, we conclude by presenting four key research directions we identified for practitioners and researchers to delineate future work.
\end{abstract}

\ccsdesc[500]{Security and privacy~Software security engineering}
\ccsdesc[300]{General and reference~Empirical studies}
\ccsdesc[500]{Software and its engineering~Agile software development}

\keywords{secure agile software engineering, secure DevOps, DevSecOps, secure continuous software engineering, security compliance, security challenges}

\maketitle

\section{Introduction and Background}\label{sec:intro}

Combining secure development practices with \acrfull{csd} workflows poses various challenges. How to cope with them has been in the scope of both researchers and practitioners alike for many years \cite{moyon:2020:mapping}.
In our perception, a certain consensus seems to be that the integration of security practices is, at least, difficult. This is corroborated by various secondary studies \cite{queslati:2015:literature,mohan:2016:secdevops, myrbakken:2017:devsecops, villamizar:2018:mapping, rajapakse:2021:challenges, ramaj:2022:holding} and more practically oriented grey literature \cite{sonatype:2019, gartner:2017} reporting on the state of the art and related challenges. 
However, existing recommendations require too often more empirical evaluations and especially acceptance by practitioners considering their aim at solving practical engineering challenges \cite{moyon:2020:mapping,Bitra:2021}.
To the best of our knowledge, while \acrshort{csd} has been adopted in various interpretations by large enterprises (e.g. scaled agile), the question of which security-specific challenges and needs are 
relevant in the context of \acrshort{csd} requires further exploration specifically in terms of relevance according to practitioners' target (e.g. software product security, secure development process, security strategy).
We consider this of particular importance to both practitioners as well as researchers developing plausible solutions to those challenges.
In this paper, we report on our study results to provide a first answer to this very question. We extend existing discussions around security challenges in agile development and, in particular, DevOps, considering two perspectives: The experiences and expert opinions of individuals, and the perspective of engineering teams. We aim to analyze how groups of practitioners perceive which challenges in their specific context, as described next. 
\subsubsection*{\textbf{Study Design in a Nutshell}} \ \\
In brief, we conducted case studies in 3 different companies operating in highly regulated environments. 

\begin{figure}[ht!]
	\centering
    \includegraphics[width=0.8\columnwidth]{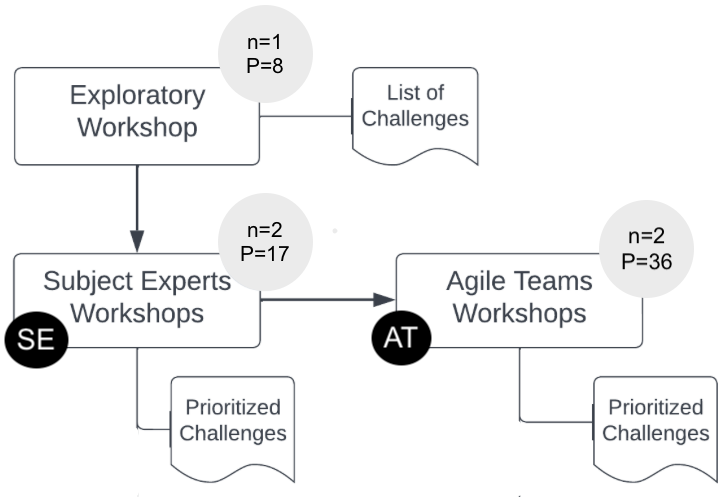}
	\small \caption[Study Protocol]{Summary of Study Methods. Circles in grey indicate: number of workshops n, and participants P.}
	\label{fig:protocol}
\end{figure}

Initially, we partnered with security experts to produce a draft list of relevant challenges in the field combining pertinent practitioners and academic sources. Later, in 2019 and 2020, we held five workshops with security and CSD practitioners to refine and validate the list of challenges (see Figure \ref{fig:protocol}). A \textit{exploratory workshop} with security experts, who consult security for agile teams, delivered the final challenges. Subsequently, individuals of two focus groups reviewed and prioritized the challenges: subject experts (SE) in \acrshort{csd} with responsibility for security, and agile team members with cross-functional roles (AT).
Our study participants chose a maximum of five challenges in \acrshort{csd}, that are of priority considering their common objective. Such an objective is either to improve the security of a specific software product or to implement strategies for secure agile development in their organization. 
The study reported here is part of our long-term investigation of our industrial partners with the goal of supporting an efficient integration of security practices in \acrshort{csd}. As we hope to foster the exchange and discussions around industrial challenges in \acrshort{csd}, we concentrate on reporting our distilled challenges and implications for further research.
\section{Industrial Challenges}
Our study yielded a list of prioritized challenges for security in \acrshort{csd} as depicted in \Cref{tab:challenges_v2}. 
The challenges are grouped into five categories: 
\begin{itemize}
    \item \textit{Continuous development}: how to perform security activities like threat modeling, secure code review, security hardening continuously, and, enabling continuous experimentation for security purposes.
    \item \textit{Value Stream}: how to make security value visible to the customers through requirements, architecture or faster feedback of security vulnerabilities.
    \item \textit{Efficiency}: how to achieve security for product and in the process with less impact in resources or lead time.
    \item \textit{Knowledge Transfer}: how to educate teams to make effective security decisions and share knowledge.
    \item \textit{CI/CD Pipelines}: how to automate security in pipelines considering its protection and compliance.
\end{itemize}
{
\small
\begin{table}[ht!]
    \centering
    \caption{Practitioner's Prioritized Challenges for Security in Continuous Software Development}
   \begin{tabular}{L{0.3cm}L{5.2cm} C{0.35cm} C{0.35cm} C{0.35cm} C{0.35cm}}
  \toprule
& \multicolumn{1}{c}{\multirow[b]{2}{*}{\textbf{CHALLENGE}}} & \multicolumn{4}{c}{\textbf{TOTAL VOTES}}\\
 \cmidrule(lr){3-6}
& & \textbf{SE W1} & \textbf{SE W2} & \multirow[t]{1}{=}{\textbf{AT W1}} & \multirow[t]{1}{=}{\textbf{AT W2}} \\
\midrule
\multicolumn{6}{c}{\textbf{Category 1: Security in Continuous Development}}\\
\midrule
\multirow[t]{1}{=}{C1} & Perform threat modeling and consistently increment / adapt it throughout sprints.  & 7 & 4& 4 & 3\\
\multirow[t]{1}{=}{C2} & Include security activities into continuous deployment. & 3 & 4 & 6 & 4\\
\multirow[t]{1}{=}{C3} & Involve security aspects into continuous experimentation or prototyping. & 0 & 1 & 3 & 4 \\
\addlinespace
\midrule
\multicolumn{6}{c}{\textbf{Category 2: Security in the Value Stream}}\\
\midrule
\multirow[t]{1}{=}{C4} & Prioritization of security requirements vs. system functionalities.  & 7 & 4 & 10 & 4\\
\multirow[t]{1}{=}{C5} & Make security architecture visible in backlog and documentation. & 1 & 8 & 8 & 0\\
\multirow[t]{1}{=}{C6} & Get security activities into the early feedback principle of DevOps.  & 0 & 9 & 1 & 2 \\
\addlinespace
\midrule
\multicolumn{6}{c}{\textbf{Category 3: Security Implementation Efficiency}}\\
\midrule
\multirow[t]{1}{=}{C7} &Balance efforts for security: improving process compliance vs. improving product quality.& 6 & 1 & 2 & 5\\
\multirow[t]{1}{=}{C8} &Involve security activities with minimal burden of lead time.& 1 & 0 & 7 & 3 \\
\addlinespace
\midrule
\multicolumn{6}{c}{\textbf{Category 4: Security Knowledge}}\\
\midrule
\multirow[t]{1}{=}{C9} &Enable security knowledge and ownership in engineering teams. & 2 & 7  & 8 & 9 \\
\multirow[t]{1}{=}{C10} & Congregate experts to interchange examples about DevOps and security. & 0 & 0 & 8 & 1 \\
\addlinespace
\midrule
\multicolumn{6}{c}{\textbf{Category 5: Security into CI/CD pipelines}}\\
\midrule
\multirow[t]{1}{=}{C11} &Effective hardening of development, test and deployment environments. & 1 & 0 & 4 & 3\\
\multirow[t]{1}{=}{C12} &Achieve efficient handling of security tool findings and involve into regular issues handling process. & 0 & 2 & 6 & 0 \\
\multirow[t]{1}{=}{C13} &Selection of security tools to run in the CI/CD pipeline. & 0 & 2 & 6 & 0\\
\multirow[t]{1}{=}{C14} &Match security compliance requirements with working pipelines. & 0 & 0 & 4 & 3 \\
\multirow[t]{1}{=}{C15} &Ensure protection of CI/CD pipelines. & 2 & 2 & 1 & 0\\
\addlinespace
\bottomrule
\end{tabular}
\label{tab:challenges_v2}
\end{table}
}

\label{sec:results}
\section{Future Research Work} \label{sec:discussion}
Considering these challenges at our industrial partner, we propose the following four major takeaways as relevant for organizations adopting \acrshort{csd} while pertaining to regulated security environments:
\begin{enumerate}
    \item \textit{Visibility and assessment of security practices maturity for CSD based on applicable security standards} allowing practitioners to break silos between engineering and security experts teams by providing transparency of the status of security practices that regulators will expect in the development process. This includes evaluating both human and automation-based security activities e.g. secure coding review by peer-reviewed and static methods. Challenges: C1,C2, C4, C5, C7, C8.
    \item \textit{Implementation of continuous security feedback-loop} referring to the effective management of security-related issues, that are identified through automatic vulnerability checking by CI/CD Pipelines. Challenges: C9, C12.
    \item \textit{Continuous automatic security compliance} assessing capabilities and limitations of automation in continuous security compliance, e.g. by providing delta analyses with particular security norms, and allowing for the development of sensible automation solutions. Challenges: C6, C7, C8, C14.
    \item \textit{Improvement of security skills in agile/DevOps teams} by training security champions and providing pertinent guidance on continuous security practices. This favors better decision-making about security risks in the team's products' continuous delivery. Challenges: C9, C10 influencing C11, C13, C15. 
\end{enumerate}

The relevance of the takeaways is further corroborated by our partner's decision to explicitly support them with dedicated internal research projects and doctoral studies (c.f. \cite{moyon:2023:refa,voggenreiter:2022:semantic,angermeir:2021:automation}).  

\subsection*{Acknowledgements}
The authors would like to thank the participants of this study, as well, as the members of the Task Forces in the industrial partners who actively support the initiative.

\bibliographystyle{ACM-Reference-Format}
\bibliography{acmart}

\end{document}